\documentclass[12pt,twoside,letterpaper]{article}

\usepackage{amssymb}
\usepackage{caption}
\usepackage{graphicx} 
\usepackage{xspace}
\usepackage{multirow}

\newcommand{\met} {\ensuremath{\mbox{$\protect \raisebox{0.3ex}{$\not$}E_T$}}\xspace}

\newcommand{\vv}{\ensuremath{\nu\bar{\nu}}\xspace}
\newcommand{\lv}{\ensuremath{\ell\nu}\xspace}
\renewcommand{\ll}{\ensuremath{\ell\ell}\xspace}
\newcommand{\bb}{\ensuremath{b\bar{b}}\xspace}
\newcommand{\hbb}{\ensuremath{H\to\bb}\xspace}
\renewcommand{\tt}{\ensuremath{t\bar{t}}\xspace}
\newcommand{\hgg}{\ensuremath{H\to\gamma\gamma}\xspace}
\newcommand{\hzz}{\ensuremath{H\to Z^{(*)}Z}\xspace}
\newcommand{\ww}{\ensuremath{W^+W^-}\xspace}
\newcommand{\hww}{\ensuremath{H\to \ww}\xspace}
\newcommand{\htautau}{\ensuremath{H\to \tau^+\tau^-}\xspace}
\newcommand{\metbb}{\ensuremath{\met+\bb}\xspace}

\newcommand{\vh}{\ensuremath{V\!H}\xspace}
\newcommand{\wh}{\ensuremath{W\!H}\xspace}
\newcommand{\zh}{\ensuremath{Z\!H}\xspace}

\newcommand{\gevcc}{\ensuremath{\mathrm{GeV}/c^2}\xspace} 
 
\newcommand{\gev}{GeV\xspace}

\begin{document}
\thispagestyle{empty}
\begin{flushright}
CDF/PUB/ELECTROWEAK/PUBLIC/10996\\
FERMILAB-CONF-13-130-E \\
Version 2.0\\
\today
\end{flushright}

\vspace{0.5in}

\begin{center}
{\bf Standard Model Higgs Boson Searches at the Tevatron}
\end{center}
\begin{center}
Kyle J.~Knoepfel\footnote{E-mail: {\texttt knoepfel@fnal.gov} }
\textsuperscript{,}\footnote{Contributed to the Proceedings of Les Rencontres de Physique de la Vall\'ee d'Aoste, March 2013, La Thuile, Italy.}\\
(\emph{On behalf of the CDF and D0 collaboration}) \\ \vspace{1.0ex}
\emph{\small Fermi National Accelerator Laboratory}
\end{center}

\begin{abstract}
Updated Standard Model Higgs boson search results from the Tevatron
experiments are presented.  We focus on the updated CDF
\metbb result, where a significant shift in observed limits is
explained.  For the Tevatron combinations, upper limits at 95\%
credibility level and best-fit values for the Higgs boson cross
section times branching ratio are presented.  We also place
constraints on the Higgs couplings to fermions and electroweak vector
bosons.  All results are consistent with the existence of a Standard
Model Higgs boson with a mass of 125~\gevcc, and with the
Standard-Model predictions associated with that assumption.
\end{abstract}

\section{Introduction}
In the context of the Standard Model (SM) of particle
physics~\cite{ref:sm}, the Higgs mechanism~\cite{ref:higgs} has been
postulated to instigate electroweak symmetry breaking, which produces
the electroweak $W^\pm$ and $Z$ bosons.  The mechanism gives rise to a
new scalar particle, the Higgs boson, which has been the object of
many experimental searches for the past few decades.

In July 2012, the ATLAS and CMS experiments separately claimed
discovery of a particle with a mass of 125~\gevcc that is consistent
with a SM Higgs boson interpretation~\cite{ref:higgsdisc}.  To
identify the new state as the SM Higgs boson, however, requires that
measurements of its couplings to fermions and electroweak bosons, as
well as measurements of its production cross-sections and decay
branching fractions are consistent with SM predictions.  To this end,
the Tevatron experiments in August 2012 jointly claimed evidence of a
new particle that decayed to a \bb pair that was consistent with the
LHC discoveries and the SM predictions~\cite{ref:tevhbb}.

Whereas the primary sensitivities of the 125-\gevcc particle at the
LHC are due to the relatively clean gluon-fusion
\hgg and \hzz modes, where both $Z$ bosons (one produced
on shell) decay to pairs of leptons, the cross sections of such modes
at the Tevatron are largely suppressed due to the lower center-of-mass
energy.  The primary sensitivities in Higgs boson searches at the
Tevatron, therefore, come mainly from the associated production modes
(\vh), where the Higgs boson is produced alongside an electroweak
boson $V$ (which represents the $W$ or $Z$), and the Higgs boson
decays via \hbb.

As the $b$-quarks are often produced at energies greater than 50~\gev
in the laboratory reference frame, they fragment into a cascade of
less energetic particles, which eventually hadronize at the scale of
$\Lambda_\mathrm{QCD}$.  Various reconstruction algorithms are
optimized to collect as much energy of these ``jets'' (and therefore
the original $b$ quark) as possible, without introducing extra energy
from same-event particles that are unassociated with the original
$b$-quark.  Although each experiment applies jet-energy corrections
which adjust the jet energies back to the quark-level quantities, the
rms of these corrections tends to be on the order of 10-20\% of the
overall scale.  If a Higgs boson that decays via \hbb exists, the
dijet invariant mass of the system is thus not able to constrain well
the mass of the Higgs boson, unlike the \hgg and \hzz searches.
Rather, the overall production rate, and other quantities must be used
to place constraints on the allowed Higgs boson mass at the Tevatron
experiments in this channel.

In this contribution, we discuss the current status of the individual
Tevatron Higgs searches (up to March 2013), as well as the Tevatron
combinations at CDF, D0, and the combined results from both
experiments.  For the combinations, in addition to excluding the Higgs
boson across a putative mass range of $100 < m_H < 200$~\gevcc, we
also present best-fit values of the cross-section times branching
ratio, and place constraints on the $HVV$ and $H\!f\!f$ couplings.  We
do not discuss fermiophobic or fourth-generation searches, which are
presented in an upcoming publication~\cite{ref:tevcombo}.

\section{Status of the Individual Tevatron Searches}

The CDF and D0 experiments have many analyses in the final stages of
presentation.  Table~\ref{tab:status} presents the \textit{current}
publication status of the Tevatron experiments that went into the
presentation at the La Thuile conference in March 2013\footnote{To
avoid confusion in citing publications, we present the publication
status of each individual search current as of submission of this
proceedings contribution, and not the status at the time of the
conference presentation.}.  Of the searches presented in
Table~\ref{tab:status}, we will focus on the CDF
\metbb Higgs search result as it is the analysis with the most
significant updates since the last conference (HCP 2012 in Kyoto,
Japan).

\begin{table}[tb]
  \caption{Status of Tevatron Higgs boson searches -- new since the HCP 2012 conference.}
  \label{tab:status}
  \begin{tabular}{cll}
    \hline\hline 
    Experiment & Search   & Status \\ \hline
      \multirow{2}{*}{CDF} & $\vh\to\metbb$         & Published in Phys. Rev. D~\cite{ref:cdfmetbb}    \\
                           & $\vh\to q\bar{q}'+\bb$ & Published in J. High Energy Physics~\cite{ref:cdfallhadronic}  \\
    \hline
      \multirow{5}{*}{D0}  & $\lv+$jets searches & Accepted by Phys. Rev. D~\cite{ref:d0lvbb} \\
                           & $H\to W^+W^-$      & Accepted by Phys. Rev. D~\cite{ref:d0hww}  \\
                           & \hgg               & Accepted by Phys. Rev. D~\cite{ref:d0hgg}  \\
	                   & $H\to W^+W^-/\tau^+\tau^-$ & Accepted by Phys. Rev. D~\cite{ref:d0hwwhtt} \\
	                   & Trilepton/same-sign $e\mu$ pairs & Accepted by Phys. Rev. D~\cite{ref:d0trilepton} \\
    \hline\hline
  \end{tabular}
\end{table}

\subsection{\metbb Higgs Search at CDF}

The $\vh\to\metbb$ Higgs search~\cite{ref:cdfmetbb} is sensitive to
the $\zh\to\vv+\bb$ and $\wh\to\lv+\bb$ processes, which contain
intrinsic missing transverse energy (\met).  In the first case, the
\met results from the undetectable \vv pair, whereas for the second
case, the \met is from the undetectable neutrino and an identified
charged lepton $\ell$.  For this search, events with one identified
charged lepton are vetoed, ensuring orthogonality of data samples with
respect to the CDF $\wh\to\lv+\bb$ search.  A minimum \met requirement
of 35~\gev is made of the event to reduce background from events with
two or more reconstructed jets produced by QCD (``QCD multijet''), but
where a significant energy imbalance occurs from jet-energy
miscorrection.  Despite this requirement, QCD multijet backgrounds
remain dominant, and multivariate algorithms are implemented to
further separate the QCD background (and other SM backgrounds) from
the Higgs boson signal.  To increase the signal-to-background ratios,
the analysis is split into tagging categories based on the probability
that the jets originated from $b$-quarks.

\begin{figure}[tb]
\includegraphics[width=0.5\textwidth]{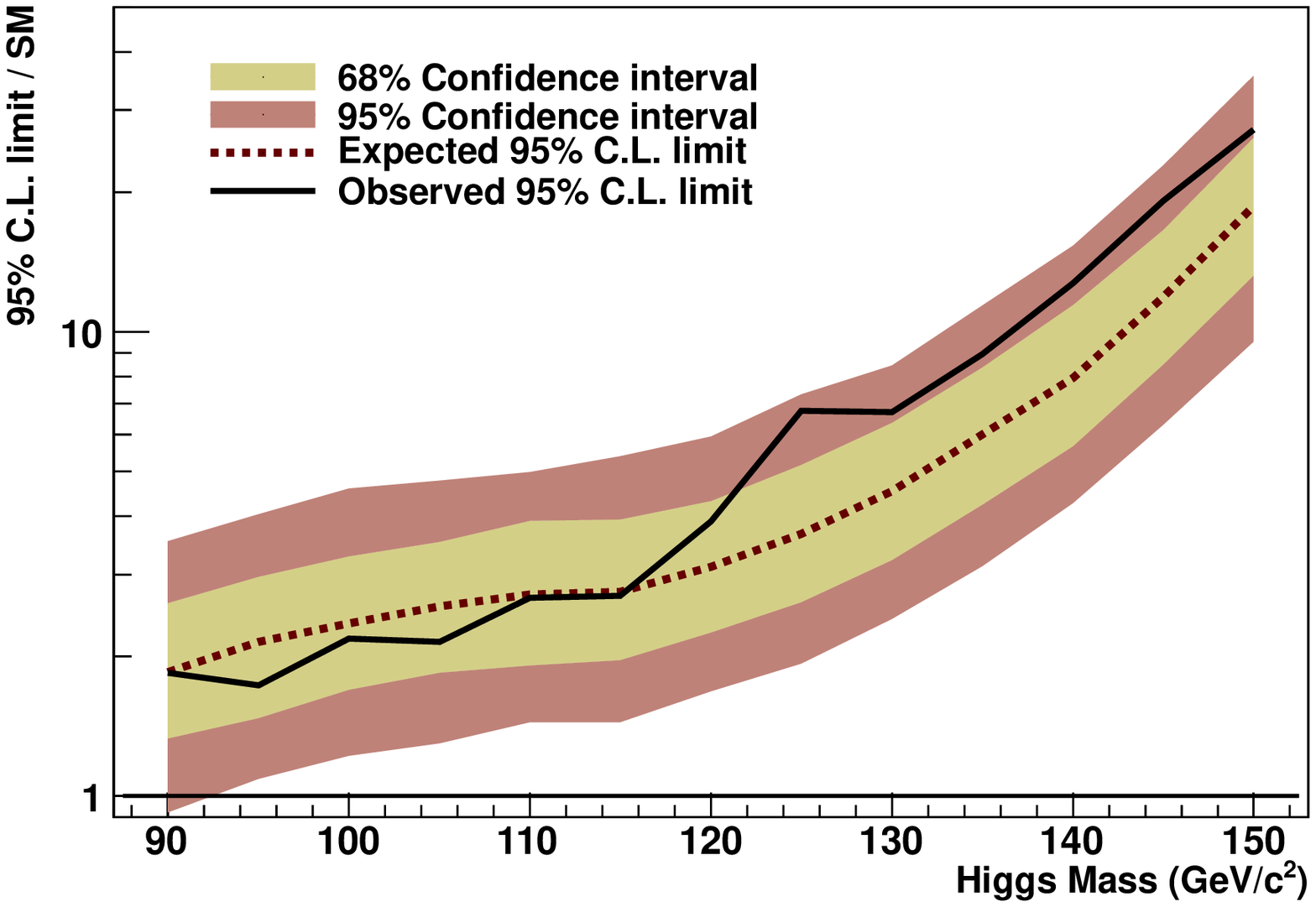} \hfill
\includegraphics[width=0.5\textwidth]{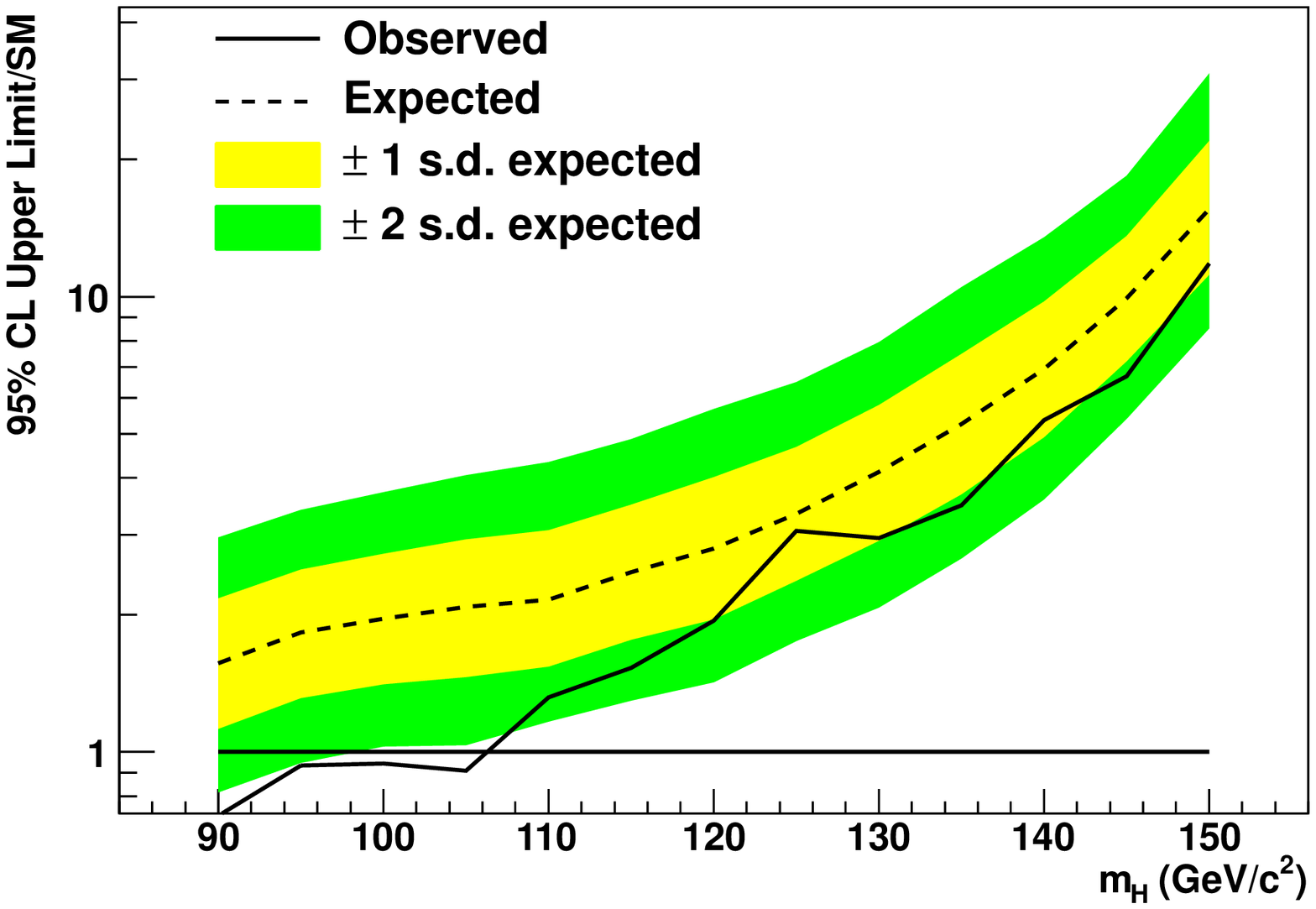}
\caption{Upper limits (95\% C.L.) on Higgs boson 
production in the \metbb channel for the (left) previous
analysis~\cite{ref:cdfmetbbold} and (right) the updated
analysis~\cite{ref:cdfmetbb}.}
\label{fig:cdfmetbblimits}
\end{figure}

To improve sensitivity to Higgs boson exclusion relative to the
previous analysis~\cite{ref:cdfmetbbold}, an updated $b$-tagging
algorithm was implemented, which was specifically optimized for \hbb
searches~\cite{ref:cdfhobit}.  Due to correlations between the new
tagging algorithm and the background modeling procedure, new QCD
background shape models and QCD-suppression multivariate algorithms
needed to be derived and retrained, respectively.
Figure~\ref{fig:cdfmetbblimits} shows the 95\% credibility level
(C.L.) limits for the previous analysis and the updated analysis.
Even though the only dominant change in analysis methodology is the
implementation of an improved $b$-tagging algorithm in the newer
analysis, a fairly significant shift in the observed limits is seen
(55\% on average), whereas a 14\% improvement is expected.

A non-negligible portion of the shift in observed limits is due to a
different treatment of systematic uncertainties between $b$-tag
categories.  However, the primary reason for the remaining change in
the shift is due to significant event migration between the
$b$-tagging categories of the previous analysis, and those of the
updated one.  A two-sided $p$-value was calculated to estimate that
the remaining shift in observed limits between both analyses was due
to statistical effects of event migration.  Accounting for the
statistical correlations between the two analyses, and the
correlations between each $m_H$ hypothesis, the probability that the
non-systematic change in observed limits is due to statistical effects
only is at the 3\%-5\% level.  As no background mismodeling was
observed in the updated analysis, and as applying the updated
treatment of systematic uncertainties to the previous analysis did not
significantly alter any of the previous results, we conclude that the
significant shift in observed limits is due primarily to statistical
effects of event migration.  For further details, see
Ref.~\cite{ref:cdfmetbb}.

\subsection{CDF Combination Considerations}

Even though the previous and updated versions of the CDF \metbb
analysis use different $b$-tagging techniques, both analyses are
robust in terms of background modeling, and in accounting for
systematic effects.  Both results are therefore interpreted as
correct, but different ways of analyzing the same Higgs boson search
channel.  For the final combination, however, CDF uses the analysis
that gives the best sensitivity to excluding the Higgs at 95\%
C.L.---thus, the updated \metbb result was used in the final CDF and
Tevatron Higgs boson combinations.

\section{Status of the Tevatron Combinations}

At the time of the conference, the final CDF Higgs boson combination
had been submitted for publication and was thus available for public
presentation~\cite{ref:cdfcombo}.  The final D0 and Tevatron
combinations were not yet public, so combinations from October 2012
and November 2012, respectively, were presented and are also shown
here.

\subsection{Upper Limits and Best-fit Values on Higgs Production}

\begin{figure}[tb]
\includegraphics[width=0.5\textwidth]{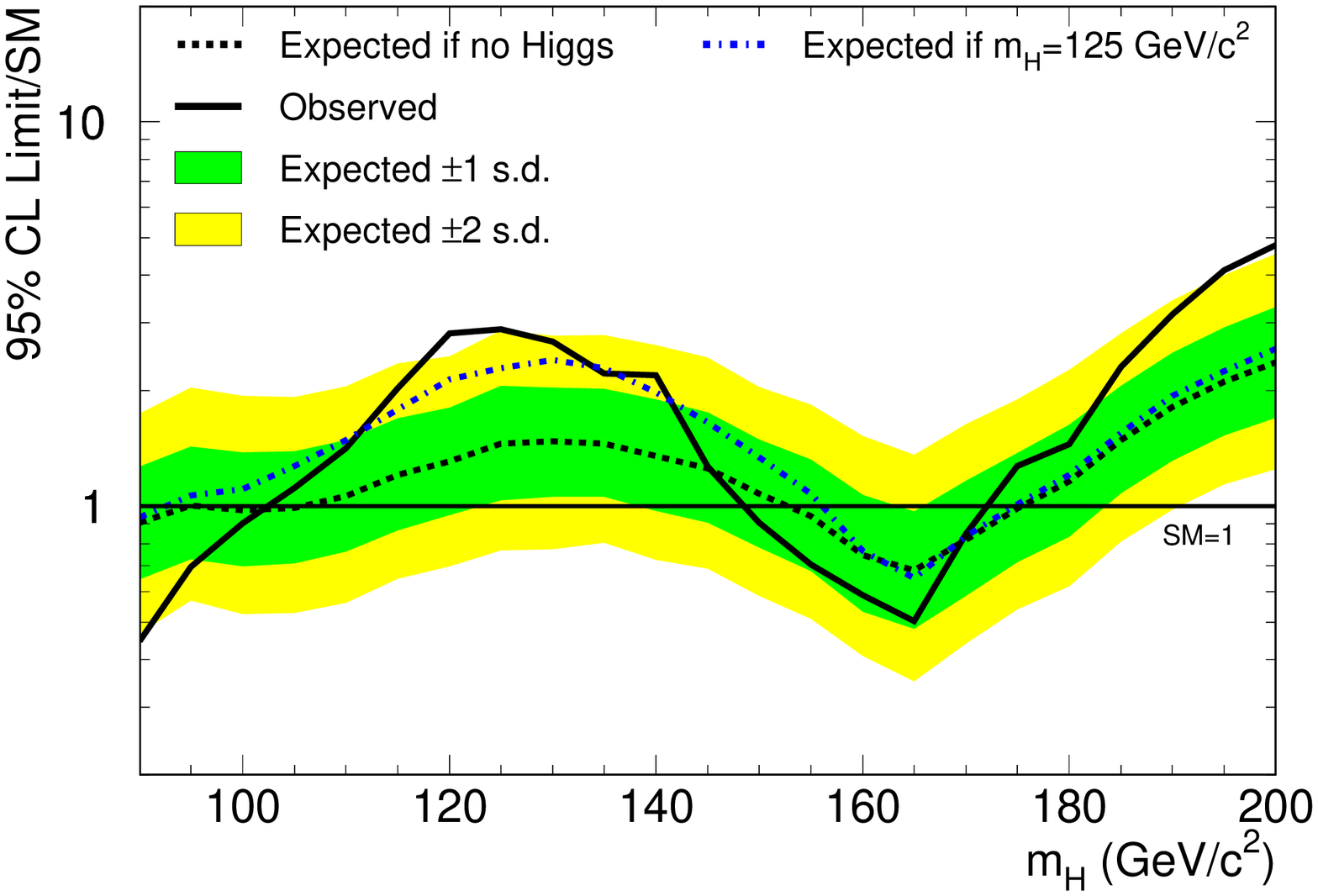} \hfill
\includegraphics[width=0.5\textwidth]{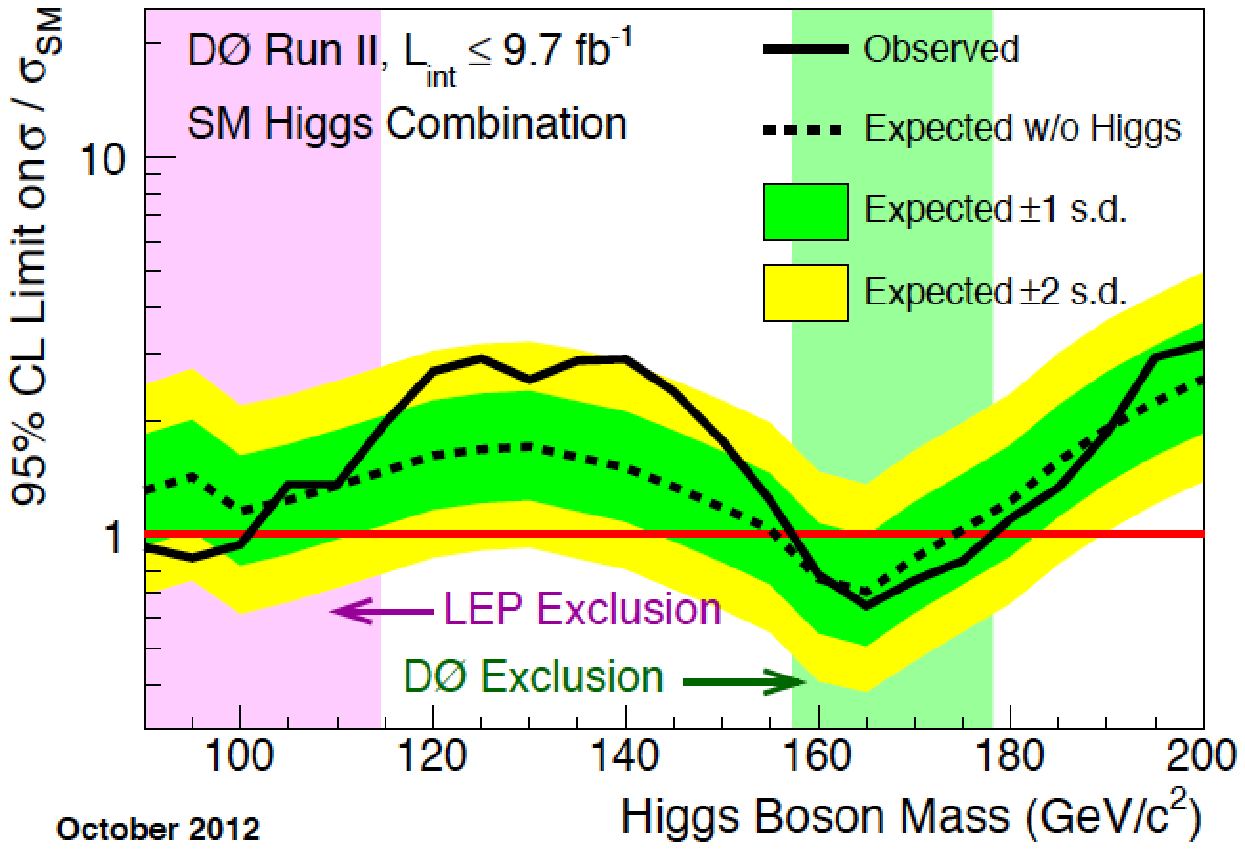}
\caption{Upper limits (95\% C.L.) on Higgs production $\sigma\times\mathcal{B}$ 
for (left) CDF and for (right) D0 in units of the SM prediction. }
\label{fig:cdfd0limits}
\end{figure}

The individual 95\% C.L. upper limits on the Higgs production cross
section times branching ratio are shown in fig.~\ref{fig:cdfd0limits}
in units of the SM prediction.  Excesses in the observed limits are
seen in both experiments in the range $100 \lesssim m_H \lesssim 150\
\gevcc$.  Higgs boson mass regions are excluded where the observed
line falls below unity.  The CDF plot also shows what one would expect
to see if there were a 125-\gevcc Higgs boson present in the data.

\begin{figure}[tb]
\includegraphics[width=0.5\textwidth]{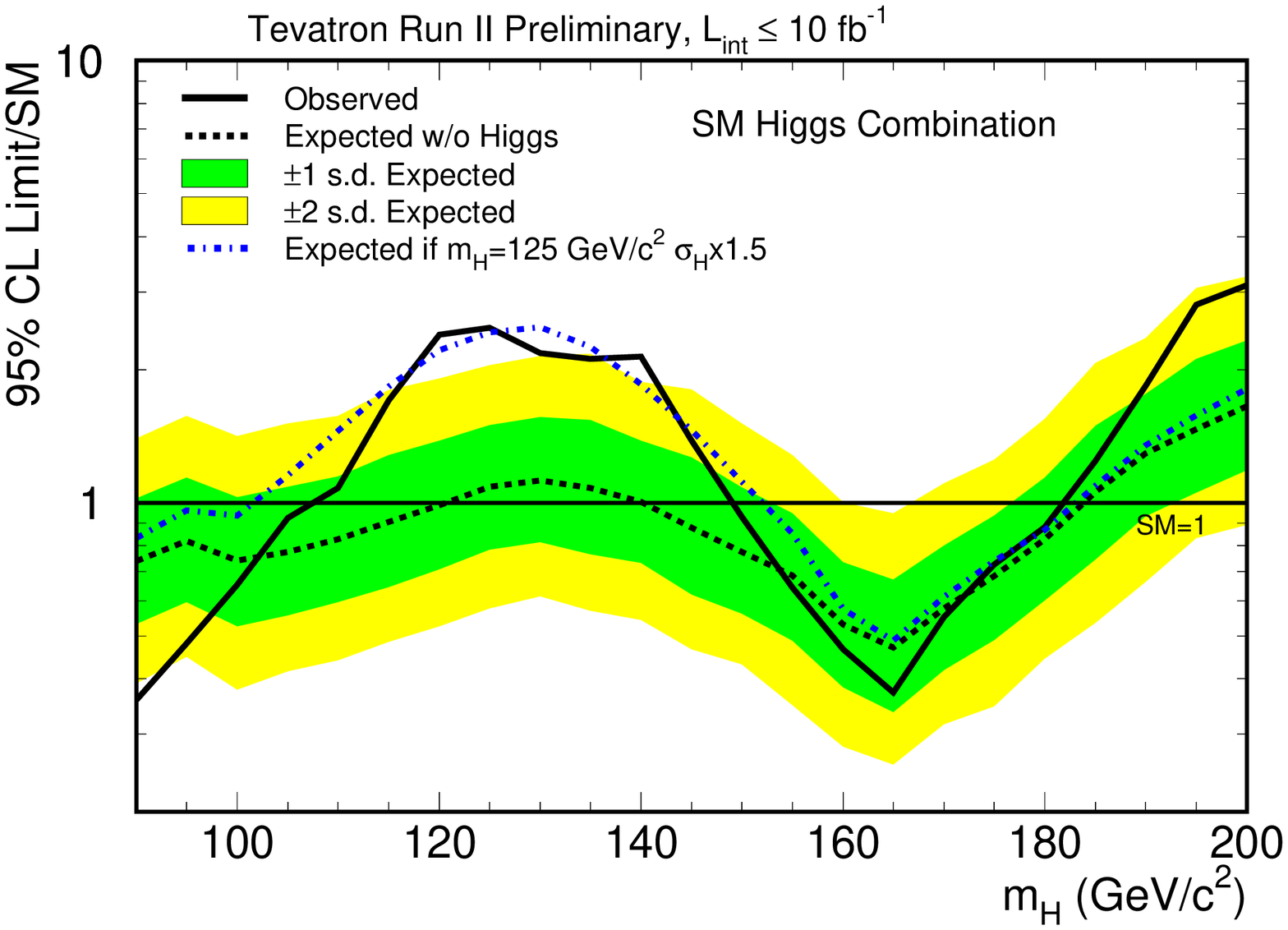} \hfill
\includegraphics[width=0.5\textwidth]{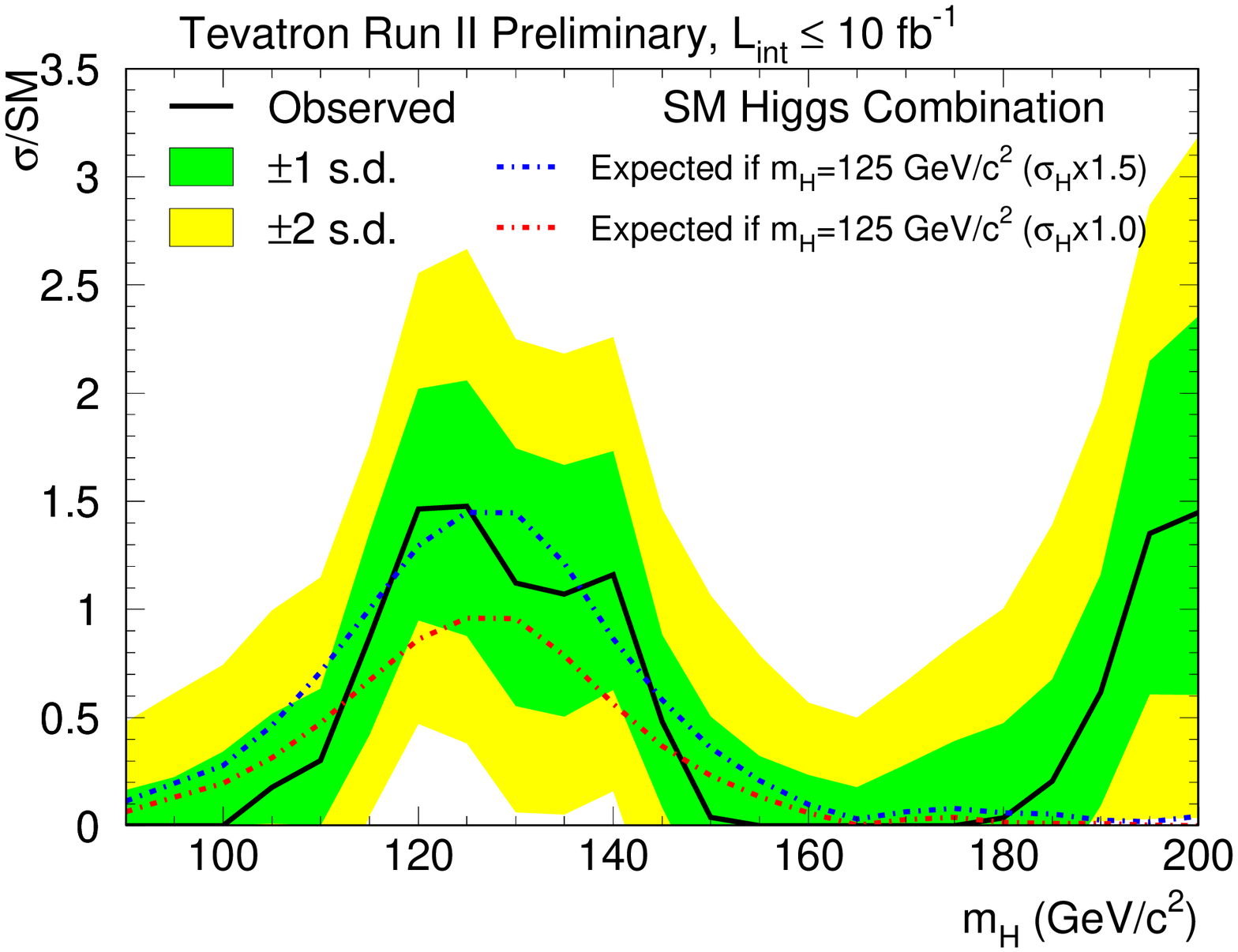}
\caption{Tevatron combination for the (left) 95\% C.L. upper limits on the Higgs 
production rate and for the (right) best-fit value of the $\sigma\times\mathcal{B}$ 
as determined from the data.  Both plots are presented in units of the SM prediction.}
\label{fig:tevcomb}
\end{figure}

\begin{figure}[tb]
\centering
\begin{minipage}{0.48\textwidth}
\includegraphics[width=\textwidth]{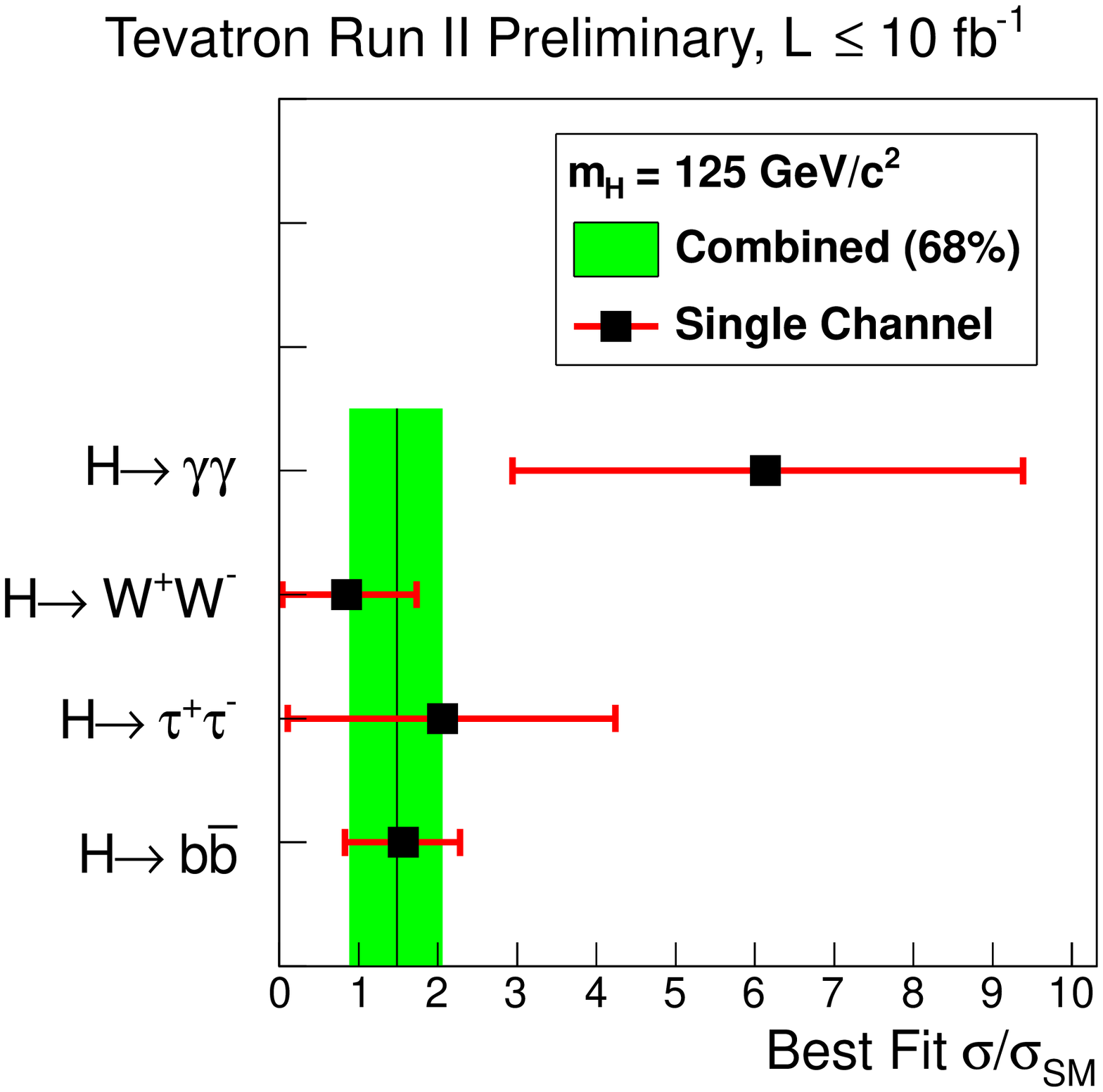} 
\end{minipage} \hfill
\begin{minipage}{0.48\textwidth}
  \renewcommand{\arraystretch}{1.3}
  \centering
     \begin{tabular}{lc} 
     \hline\hline
     \multicolumn{2}{c}{Preliminary Results (Nov. 2012)} \\
  Process & Best-fit $\sigma\times\mathcal{B}$/SM \\ \hline
  \hww    & $0.88^{+0.88}_{-0.81}$ \\
  \hbb    & $1.56^{+0.72}_{-0.73}$ \\
  \hgg    & $6.13^{+3.25}_{-3.19}$ \\
  \htautau& $2.12^{+2.25}_{-2.12}$ \\ 
  \hline
  Combined& $1.48^{+0.58}_{-0.60}$ \\
  \hline\hline
  \end{tabular}
\end{minipage}
\caption{Best-fit values for Higgs boson cross section and times branching ratio 
for individual search channels, as well as for the combined result.
The table at the right is the numerical form of the plot on the left.}
\label{fig:xsec125} 
\end{figure}

The CDF-D0 combined (Tevatron) upper limits and best-fit production
rates are shown in fig.~\ref{fig:tevcomb}, where correlated
uncertainties between both experiments have been taken into account.
Both plots show the expected shape of the data if a 125-\gevcc Higgs
boson were present in the data, produced at 50\% greater rate than is
predicted in the SM.  The right plots also shows the best-fit value
for the Higgs boson production produced at the nominal SM prediction.
As can be seen, the Tevatron data prefer a scenario that assumes the
presence of a Higgs boson instead of the non-Higgs boson hypothesis
(black, dashed line).  One can take a slice of the right plot in
fig.~\ref{fig:tevcomb} for $m_H = 125\ \gevcc$ and decompose the
best-fit $\sigma/\sigma_\mathrm{SM}$ into the individual search
channels.  This is shown in fig.~\ref{fig:xsec125}.  The combined and
individual-channel best-fit results are consistent with the SM
predictions to within one standard deviation, with the exception of the
\hgg search, which exceeds it by roughly 1.5 standard deviations.

\subsection{Constraints on Higgs Couplings}

In addition to deriving limits and extracting best-fit values on the
Higgs boson cross section times branching ratio, the Tevatron
experiments also place constraints on the Higgs couplings to fermions
and the electroweak vector bosons.  This is done by introducing
coefficients $\kappa_i$ that scale the $Hi\bar{i}$ SM couplings, where
$i = f$ (fermions), $i = Z$, $i=W^{\pm}$, or $i=V$ when no distinction
is made between the electroweak vector bosons.  The SM couplings are
obtained when $\kappa_i = 1$.  At the Tevatron, the most
sensitive-to-exclusion search channels have $\sigma\times\mathcal{B}$
expressions that are mostly proportional to the product
$\kappa_f\kappa_V$.  However, the $\sigma\times\mathcal{B}$
expressions of the less sensitive channels $\tt H \to \tt+\bb$ and $\vh
\to V + \ww $ are proportional to $\kappa_f^2$ and $\kappa_V^2$, respectively.  
Search channels that do not dominate the exclusion sensitivity can
therefore provide sensitivity to constraining the Higgs couplings.

\begin{figure}
\includegraphics[width=0.5\textwidth]{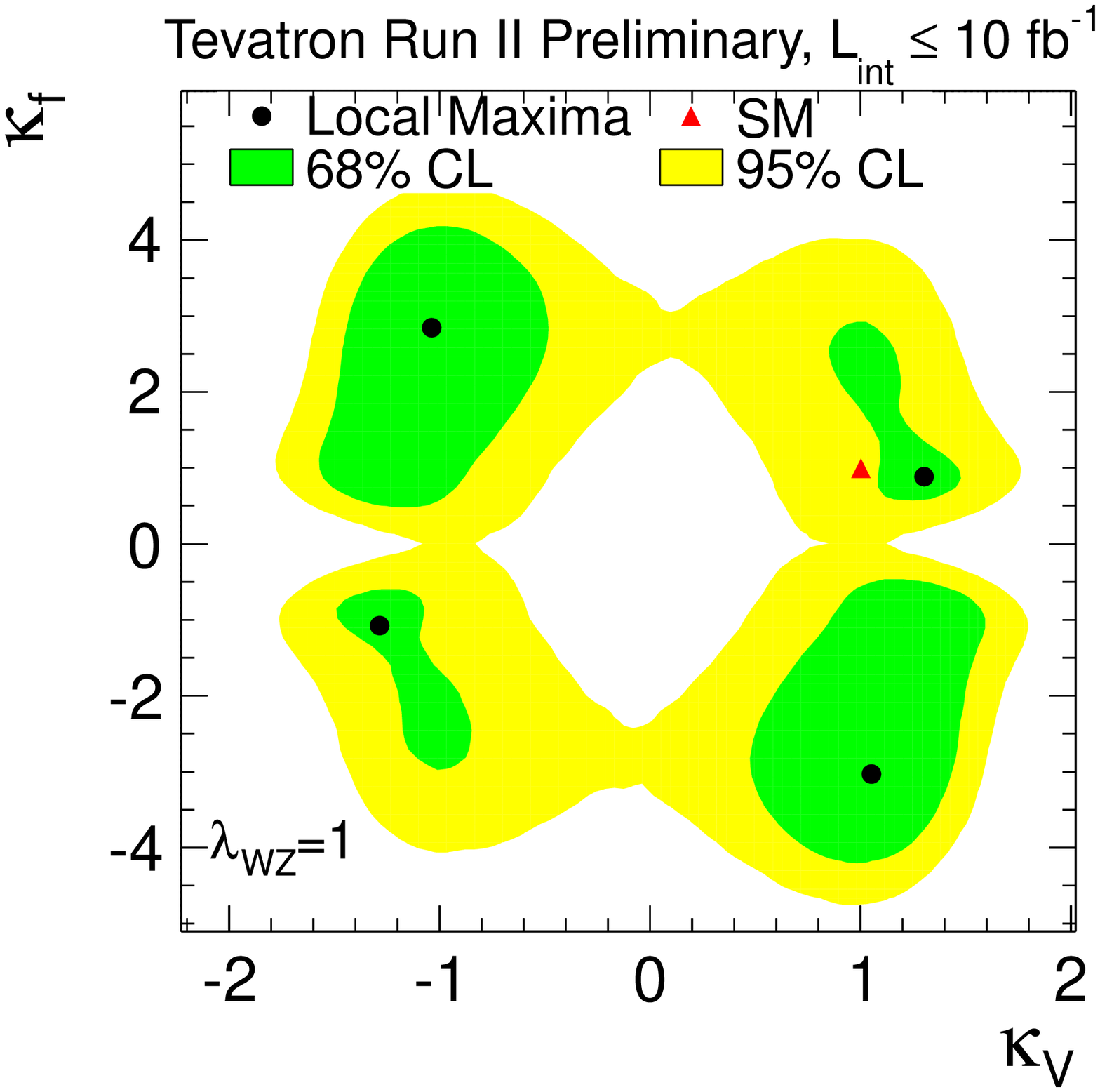} \hfill
\includegraphics[width=0.5\textwidth]{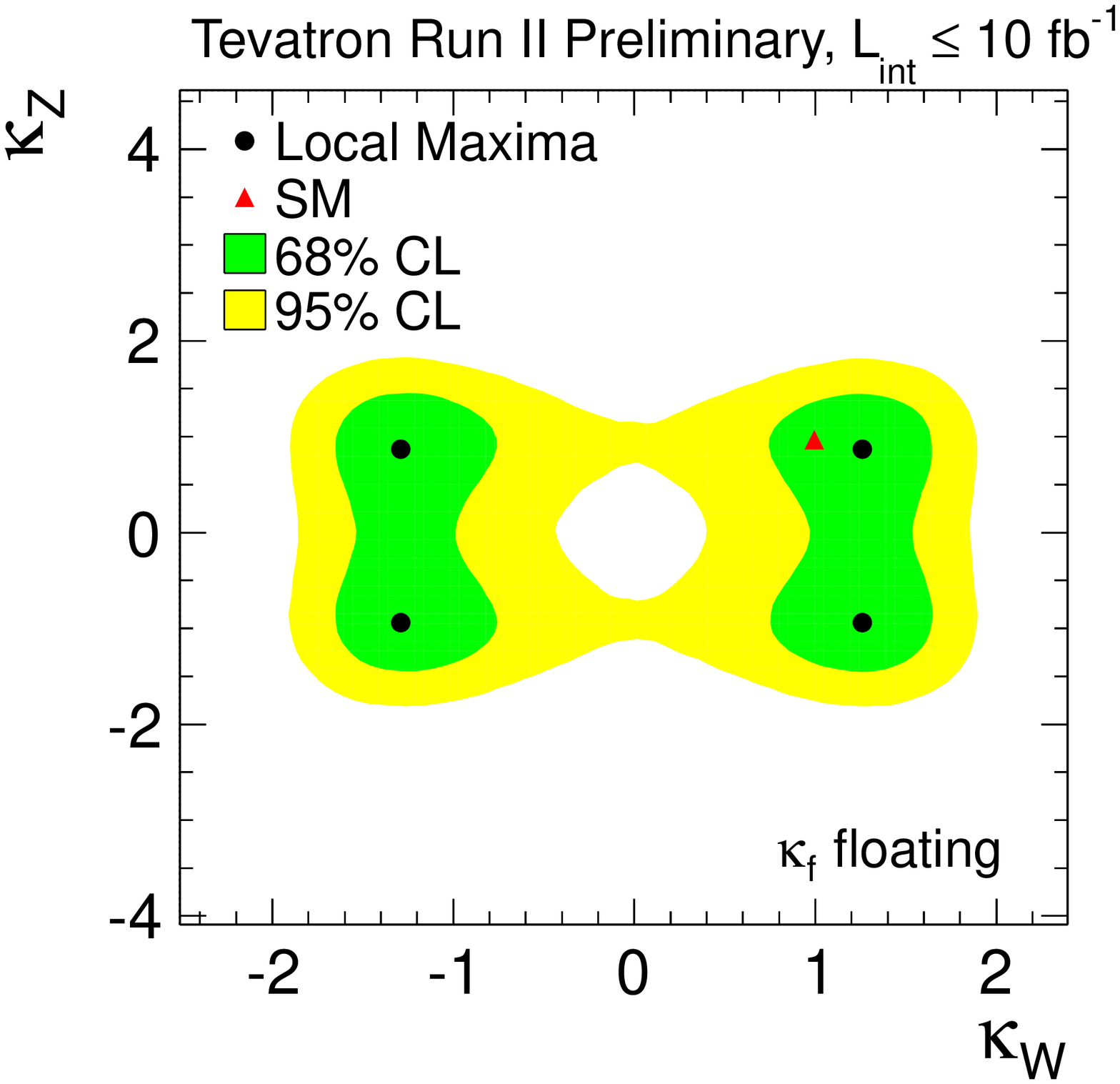}
\caption{Two-dimensional posterior probability densities for constraining 
(left) $\kappa_f$ \textit{vs}. $\kappa_V$, and (right) $\kappa_Z$
\textit{vs}. $\kappa_W$.  For the left plot $\kappa_W/\kappa_Z$ is assumed to 
be unity, whereas for the right plot, $\kappa_f$ is allowed to vary.}
\label{fig:kappa}
\end{figure}

Figure~\ref{fig:kappa} shows two-dimensional posterior probability
densities for constraining $\kappa_f$ \textit{vs}. $\kappa_V$, and $\kappa_Z$
\textit{vs}. $\kappa_W$.  Due to an interference term in the \hgg $\sigma\times\mathcal{B}$
expression, the excess in the \hgg search leads to a slight preference
for solutions in the second and fourth quadrants in the left plot of
the fig.~\ref{fig:kappa}.  The SM prediction, however, is consistent
with the Tevatron data just outside of one standard deviation.  The
right plot of fig.~\ref{fig:kappa} tests for custodial symmetry, which
in the SM guarantees $\kappa_Z = \kappa_W = 1$.  The Tevatron data are
consistent with this prediction to well within one standard deviation.

\section{Conclusions}

Whereas the results of the individual search channels have not changed
greatly since November 2012, the updated CDF \metbb
analysis~\cite{ref:cdfmetbb} has extensively studied the change in
observed limits since the publication of the previous
result~\cite{ref:cdfmetbbold}.  The large change in observed limits is
due to statistical effects of event migration by switching to an
improved $b$-tagging algorithm.  

We have presented Tevatron combinations of upper limits (95\% C.L.)
and best-fit values of the Higgs boson cross section times branching
ratio.  We see an excess in data that is consistent with a 125-\gevcc
Higgs boson interpretation.  In addition, we place constraints on the
Higgs couplings to fermions and electroweak vector bosons, the results
of which are largely consistent with SM predictions.

\appendix
\section*{Addendum}
Since the conference, additional analyses have been submitted and
accepted for publication from the D0 collaboration: the
above-mentioned full Tevatron combination~\cite{ref:tevcombo}, the
$\zh\to\ll+\bb$ search~\cite{ref:d0llbb} and the full D0
combination~\cite{ref:d0combo}.

\section*{Acknowledgments}
The author would like to thank the organizers for an enjoyable
experience at the Les Rencontres de Physique de la Vall\'ee d'Aoste
conference, and the support of Fermilab and the CDF experiment.

\end{document}